\documentstyle[aps]{revtex}

\begin{document}
\draft
\input{epsf}


\title{ Pseudorapidity Distribution of Charged Particles in
        $\overline{p}p$ Collisions at $\sqrt{s}=630\rm\,GeV$ }

\author{%
 R.~Harr,$^5$
 C.~Liapis,$^6$
 P.~Karchin,$^5$
 C.~Biino,$^4$
 S.~Erhan,$^2$
 W.~Hofmann,$^1$
 P.~Kreuzer,$^2$
 D.~Lynn,$^{2*}$
 M.~Medinnis,$^2$ 
 S.~Palestini,$^4$
 L.~Pesando,$^4$
 M.~Punturo,$^3$
 P.~Schlein,$^2$
 B.~Wilkens,$^{1\dag}$
 J.~Zweizig$^2$ }

\address{
 (1) Max-Planck-Institut f\"ur Kernphysik, Heidelberg, Germany \\
 (2) University of California, Los Angeles, U.S.A. \\
 (3) University of Perugia and INFN, Italy \\
 (4) University of Torino and INFN, Italy \\
 (5) Wayne State University, Detroit, Michigan 48201 U.S.A. \\
 (6) Yale University, New Haven, Connecticut, U.S.A. }

\date{\today}

\maketitle

\begin{abstract}
Using a silicon vertex detector, we measure the charged particle
pseudorapidity distribution over
the range 1.5 to 5.5 using data collected from 
$p\overline{p}$ collisions at $\sqrt{s}=630\rm\,GeV$.
With a data sample of $3\times10^6$ events, we deduce a
result with an overall normalization uncertainty of $5\%$,
and typical bin to bin errors of a few percent.
We compare our result to the measurement of UA5, and the distribution
generated by the Lund Monte Carlo with default settings.
This is only the second measurement at this level of
precision, and only the second measurement
for pseudorapidity greater than 3.
\end{abstract}



The mean angular density of particles produced in high energy
hadron-hadron collisions  is a fundamental property of the strong
interaction.
Multi-particle production is inherently a non-perturbative process,
it is currently best treated theoretically using Monte
Carlo models implemented in computer programs such as Herwig \cite{herwig},
Isajet \cite{isajet} and PYTHIA \cite{pythia}.
Besides its fundamental importance,  the angular particle density is a
primary consideration in experiment design.
Estimates of trigger rates and occupancies for future LHC detectors
depend on extrapolations of our knowledge of particle production
from energies of less than $2\rm\,TeV$ to $14\rm\,TeV$.
Future experiments to search for the quark-gluon plasma will require
an accurate knowledge of the pseudorapidity distribution of single
nucleon collisions to compare with heavy ion results.

We report on the 
angular density of charged particles produced in 
$630\rm\,GeV$ $p\overline{p}$ collisions at the
CERN S$p\overline{p}$S collider.
A convenient way to characterize the angular density is through the
pseudorapidity distribution, $dN_{ch}/d\eta$, where
$N_{ch}$ is the number of charged primary particles per collision, and
$\eta=-\ln\tan(\theta/2)$,
$\theta$ being the polar angle from the collision axis in the center
of mass.
Previous measurements of $dN_{ch}/d\eta$ in this energy range
have been reported
by the UA5 \cite{UA5}, UA1 \cite{UA1}, and CDF \cite{CDF}\ experiments.
Of these, only the UA5 result is of high precision.
It is also the only measurement covering $\mid\eta\mid>3$.

The data reported here are from the test of a
forward geometry silicon micro-vertex
detector proposed as part of a hadronic B-physics experiment (P238)
\cite{P238}.
The geometry of the detector and the trigger play an important role in
this analysis.
As shown in Fig.~\ref{fig:detector}, the detector consists of 6
tracking planes. 
Each plane is composed of 4 quadrants, and
each quadrant contains 2 silicon micro-strip detectors (MSD's)
oriented orthogonally.
One measures the $x$ coordinate and the other measures the $y$
coordinate.
The coordinate system of the experiment is oriented with $z$ along  
the direction of the proton beam, $x$ horizontal, and $y$ vertical.
The MSD's have a diode pitch of $25\rm\,\mu m$ with every second diode
readout through a capacitively coupled electrode.
The MSD's have an active area of $4.48\rm\,cm\times4.48\,cm$.

The overall length of the detector is less than $20\rm\,cm$.
Due to its small size relative to the
longitudinal extent of the luminous region ($\sigma\approx12\rm\,cm$),
we reconstruct particles with $1<|\eta|<6$, depending on the
longitudinal position of the primary interaction (see Fig.~\ref{fig:acc}).
However, the detector covers about 3 units of $\eta$
for any given position of the primary interaction.
The positions of the MSD's are determined from analysis of the data
to a precision of a few microns in $x$ and $y$ and about
$100\rm\,\mu m$ in $z$. \cite{align}
First, the MSD's within a quadrant are aligned by optimizing track
residuals.
Second, the quadrants are aligned to each other by optimizing
vertex residuals.
In the second step, the $z$-axis of the coordinate system is forced
to coincide with the axis of the luminous region.
The procedure is described in detail in Ref.~\cite{align}.
After alignment, we are able to determine the spatial
distribution of primary interactions to high accuaracy.

About 5 million events were recorded to tape with a minimum bias trigger.
The trigger consists of 2 scintillation counters placed roughly
$\pm 3\rm\,m$ from the detector.
A coincidence of the 2 counters in time with the beam crossing
forms the trigger.

Event reconstruction begins by forming clusters from adjacent fired strips.
Tracks are formed by grouping clusters on different planes that lie within
a road of $50\rm\,\mu m$ half width \cite{P238}.
The $x$-$z$ and $y$-$z$ track projections (referred to as $x$ view
and $y$ view tracks) are reconstructed independently.
The two views are coupled for the purposes of finding a vertex,
since the $z$ coordinate of the vertex is common to all the track
projections in an event.
Vertex finding begins by determining
the intercepts of the track projections with the beamline.
A cluster of intercepts from at least 2 $x$ and 2 $y$ view
tracks within $1\rm\,cm$ forms the seed for a vertex.
We assign all tracks passing within $600\rm\,\mu m$ of the position
of the seed to the vertex.
The process is repeated to search for other vertices
using the tracks not previously assigned to a vertex.

We derive $dN_{ch}/d\eta$ from the data in two ways:
from the $x$ and $y$ view tracks, and
from a subset of track projections that are matched in $x$ and $y$ to
yield a 3-dimensional track \cite{Bernd}.
Only events with at least one reconstructed vertex are selected for
inclusion in the $dN_{ch}/d\eta$ analysis.
Only tracks assigned to a vertex are used in the analysis, removing
most spurious tracks as well as
many tracks from secondary decays (primarily $K_s^0$'s and $\Lambda$'s).

Letting $u$ stand for $x$ or $y$, we define the $\eta$ projections
$\eta_{u} = -\ln\tan(\theta_{u}/2)$, where $\theta_{u}$ is the projected
angle of the track with respect to the beam axis.
Using $u$ view tracks, we form a binned $dN_{ch}/d\eta_u$ distribution
whose $i^{th}$ bin covers the interval $\eta_u^i$ to $\eta_u^i+\Delta\eta_u$.
The raw $dN_{ch}/d\eta_u$ distribution from the data is corrected, using
the Monte Carlo simulation, for any remaining spurious or secondary
decay tracks and tracks from single diffractive interactions.
The corrections for spurious tracks and tracks from secondary decays
are each about $2\%$.
The single diffractive interaction correction is about $0.5\%$.
These corrections have roughly the same shape over the measurement
interval as the corresponding $dN_{ch}/d\eta_u$ distribution.

The simulation is used to unfold $dN_{ch}/d\eta$ from 
$dN_{ch}/d\eta_u$ by minimizing
\begin{equation}
 \chi^2 = \left({\bf A}_{u}\vec{e}-\vec{g}\right)^{T}
  {\bf V}_{\!g}^{-1}\left({\bf A}_{u}\vec{e}-\vec{g}\right) \ ,
 \label{eq:unfold}
\end{equation}
where $\vec{e}$ represents the unnormalized result vector for 
$dN_{ch}/d\eta$, and
$\vec{g}$ represents the unnormalized vector of measurements.
The matrix ${\bf A}_u$ is determined from the simulation and accounts
for the effects of acceptance and
smearing as well as projecting $dN_{ch}/d\eta$.
The $(ij)^{th}$ element of ${\bf A}_u$ is 
the fraction of charged primary particles generated with
$\eta$ in bin $j$ that are reconstructed and have $\eta_u$ lying
in bin $i$.
Simulated events are reconstructed with the same code as actual data.
The covariance matrix, ${\bf V}_{\!g}$, accounts for the finite
statistics of the data and simulation.

The second method determines $dN_{ch}/d\eta$ directly (without unfolding)
using tracks reconstructed in 3-dimensions.
We match $x$ and $y$ view tracks whenever they traverse a unique
combination of MSD's.
The probability that a match can be made depends on the overall
distribution of tracks in the event and is determined by simulation.
An additional source of background arises from incorrect matchings.
We simulate and correct for this background as well as
spurious and secondary tracks.
Eq.~\ref{eq:unfold}\ is used to determine $dN_{ch}/d\eta$ 
from the matched tracks,
where ${\bf A}$ contains the effects of acceptance, smearing, and the
probability of matching.

The quantity $\vec{e}$ which minimizes Eq.~\ref{eq:unfold}\ is proportional
to $dN_{ch}/d\eta$,
\begin{equation}
  \frac{dN_{ch}}{d\eta} = \frac{N_{MC}}{N_{gen}N_{data}\Delta\eta}\vec{e} \ ,
\end{equation}
where $N_{data}$ is the number of vertices reconstructed in the data,
$N_{MC}$ is the number reconstructed in the simulation,
$N_{gen}$ is the number of generated interactions,
and $\Delta\eta=0.25$ is the bin width.

A good simulation is needed to make the corrections for
acceptance and reconstruction efficiency.
We use a detailed detector simulation based
on GEANT~3.15, coupled to PYTHIA~5.7 and JETSET~7.4 for event generation.
All the standard features of GEANT are used:
delta-ray production, Moli\`ere model of Coulomb scattering,
decays of long lived particles, photon conversions,
hadronic interactions, etc.
The simulation of the MSD's is done at the individual diode level,
and includes charge sharing between diodes and detector noise.
The positions of the MSD's used in the simulation are determined from 
the data.
During analysis, signals on non-functioning channels are suppressed,
and random detector inefficiencies are introduced, according to the
detector performance for the data set being simulated.

Events are generated with a spatial distribution of primary interactions and
pile-up rate determined from the data \cite{pileup}. 
The charged particle multiplicity distribution is generated according
to the measurement of UA5 \cite{UA5}.
To simulate the UA5 distribution, we first enable {\em multilple
  interactions} in PYTHIA, then weight the events.
The scintillation counters comprising the trigger and
the section of beampipe between the vertex detector and the counters
are included in the simulation \cite{Liapis}.
The trigger is simulated by requiring a minimum
energy deposition in each counter for the event to be triggered.
Due to the trigger, the probability for finding a track has a maximum
around $|\eta|$ of 3.

The results for approximately $3\times10^6$ triggers
are shown in Fig.~\ref{fig:allruns}.
The results are shown for $x$ view, $y$ view, and matched tracks.
The agreement is good between the $x$ and $y$ view points up to $|\eta|=4$.
Furthermore, the shapes of all 3 measurements are quite similar.
The differences in the shapes are used to determine the errors,
as described below.
The results of Fig.~\ref{fig:allruns}\ remain consistent when subdivided
by direction of track ($\pm z$), detector quadrant, or number of
detector planes each track traverses \cite{memo}.

We calculate the final result as the average of the unmatched
(x-view and y-view), and matched track distributions.
The error has 2 contributions: an overall normalization error of 
5\% (half of the 10\% difference between matched and
unmatched), and
individual bin errors relating to the shape of the distribution.
While the normalization error is entirely systematic, the shape error
has both systematic and statistical components.
The shape error is determined from $dN_{ch}/d\eta$ results
obtained from 6 independent subsets of the data.
The subsets consist of data collected during particular running periods,
and contain comparable statistics.
The average of the matched and unmatched results is found for each subset.
The deviations of these averages from the final result are displayed
in Fig.~\ref{fig:differences}.
Treating each as statistically equivalent, the shape error is
estimated from the RMS spread of the points.
Note that the excellent agreement of these results is obtained with
no adjustments to the normalization, attesting to the consistency of
the result.
Half the difference between the $x$ and $y$ results is added in quadrature
to obtain the total shape error.

We find a number of other possible systematic errors to be negligible
in comparison to the systematic errors cited above \cite{Liapis,memo}.
These include uncertainties due to
the detector efficiencies,
the locations of the MSD's,
the positions of the trigger scintillators,
the charged multiplicity distribution,
the $dN_{ch}/d\eta$ distribution generated in the simulation,
the single diffraction cross section \cite{UA4},
and the requirements for matched tracks.
We also calculated $dN_{ch}/d\eta$ using subsets of the data which
might be sensitive to systematic errors.
The subsets consist of: tracks intersecting a specific number of
planes, sensitive to
  errors in detector efficiencies and misalignments;
tracks in a particular detector quadrant, sensitive to the
  relative alignment of MSD's within a quadrant; and
tracks with positive (negative) values of $\eta$, sensitive to
  the longitudinal distribution of interactions.
From these data subsets we find no indication of additional systematic
effects.

The $dN_{ch}/d\eta$ values and the shape errors are 
listed in Table~\ref{tab:result}.
Fig.~\ref{fig:result}\ displays our result along with the
UA5 result \cite{UA5}\ and the distribution generated by PYTHIA,
unweighted for the purposes of comparison.
The prediction of PYTHIA agrees quite well 
with our result in shape,
but is low in normalization.
Our agreement with the UA5 measurement is reasonable, however
there is an indication that our measurement falls off less rapidly at high
$|\eta|$.

In summary, we have made the second high precision measurement of
$dN_{ch}/d\eta$ at a high energy hadron collider.
The most significant systematic error is the $5\%$ normalization error
arising from the difference between the matched and unmatched track results.
Errors related to the shape of the distribution are at the level of
a few percent.
We find good agreement with the shape prediction of the Lund Monte Carlo.
We find reasonable agreement with previous measurements, though
there is an indication that our measurement falls off less rapidly at high
$|\eta|$ compared with UA5.

\acknowledgements

This work was supported by the United States Department of Energy and
National Science Foundation, the Italian Istituto Nazionale di Fisica
Nucleare, and Wayne State University.
We are grateful to the Yale HEP groups for sharing their computer
resources during this analysis.
We thank the UA5 collaboration for providing us with the numerical
values of their result.

\begin{table}
 \caption{The values of $dN_{ch}/d\eta$ and their shape errors.
  The 5\% normalization error is not included.
 \label{tab:result}}
 \begin{tabular}{ @{\hspace{.5in}} c c c @{\hspace{.6in}} }
   $\eta$ range & $dN_{ch}/d\eta$ & error \\ \hline
  1.50-1.75 & 3.30 & 0.16 \\
  1.75-2.00 & 3.23 & 0.11 \\
  2.00-2.25 & 3.30 & 0.07 \\
  2.25-2.50 & 3.25 & 0.04 \\
  2.50-2.75 & 3.16 & 0.04 \\
  2.75-3.00 & 3.02 & 0.03 \\
  3.00-3.25 & 2.88 & 0.02 \\
  3.25-3.50 & 2.71 & 0.03 \\
  3.50-3.75 & 2.57 & 0.04 \\
  3.75-4.00 & 2.40 & 0.05 \\
  4.00-4.25 & 2.24 & 0.09 \\
  4.25-4.50 & 2.07 & 0.12 \\
  4.50-4.75 & 1.84 & 0.16 \\
  4.75-5.00 & 1.68 & 0.20 \\
  5.00-5.25 & 1.45 & 0.32 \\
  5.25-5.50 & 1.19 & 0.47 \\
 \end{tabular}
\end{table}

\newlength{\figwidth}
\setlength{\figwidth}{\textwidth}

\begin{figure}
 \epsfclipon \epsfxsize=\figwidth
 \epsffile[25 150 525 650]{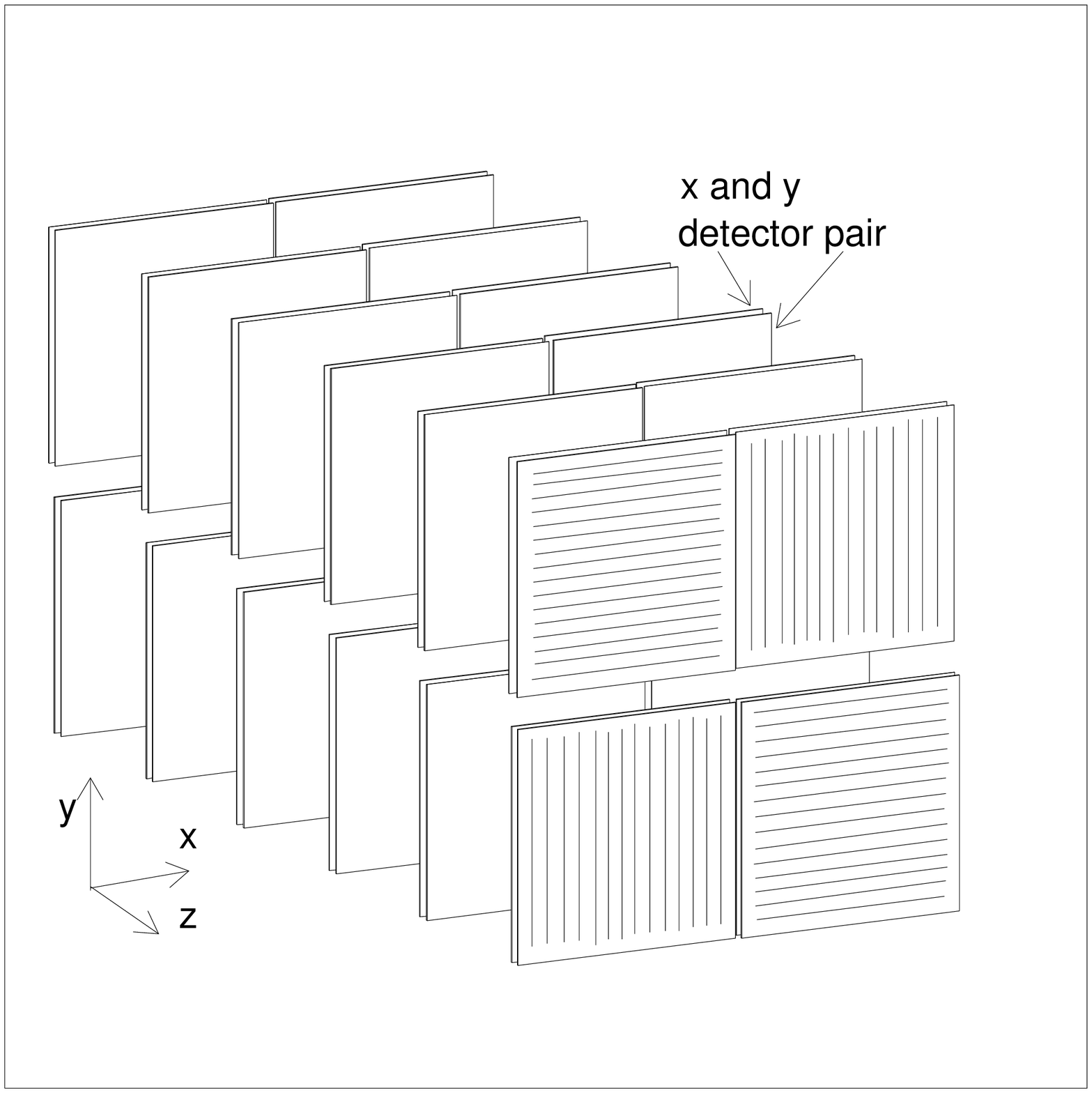}
 \caption{Isometric view of the layout of the MSD's of the P238 test.
  The $\overline{p}$ and $p$ beams pass through the gap between the
  upper and lower sets of MSD's.
  The horizontal and vertical lines represent the directions of the
  strips on the MSD's.
  Each MSD is $4.5\times4.5\rm\,cm$ and planes are separated by
  $3.8\rm\,cm$.
 \label{fig:detector} }
\end{figure}

\begin{figure}
 \epsfclipon \epsfxsize=\figwidth
 \epsffile[0 140 567 420]{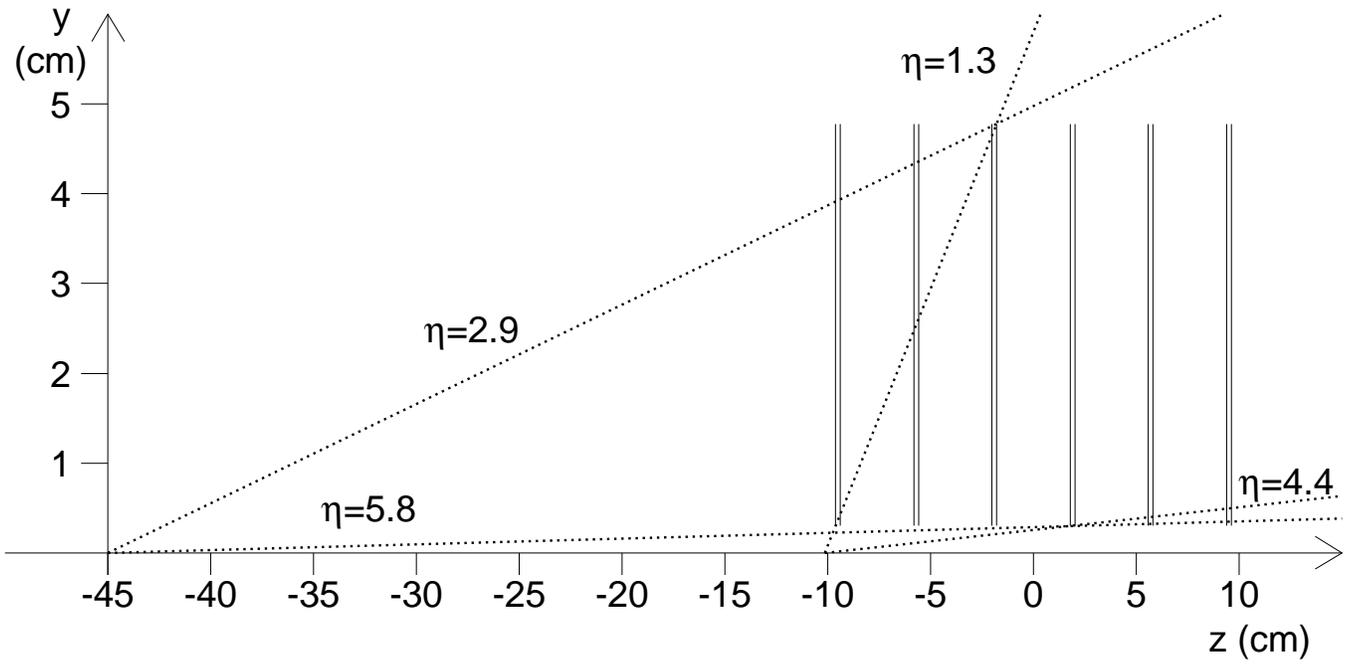}
 \caption{Side view of the upper half of the detector showing the acceptance
  limits for tracks from vertices at $z=-45\rm\,cm$ and $-10\rm\,cm$.
  The luminous region is centered around $z=-8\rm\,cm$ with
  an rms of about $13\rm\,cm$ in $z$.
  For this analysis, we use events produced within $-45\rm\,cm<z<30\rm\,cm$. 
  Although the acceptance limits change by about 1.5 units of $\eta$
  between the 2 positions, the total coverage is about 3 units of $\eta$
  for each.
 \label{fig:acc} }
\end{figure}

\begin{figure}
 \epsfclipon \epsfxsize=\figwidth
 \epsffile{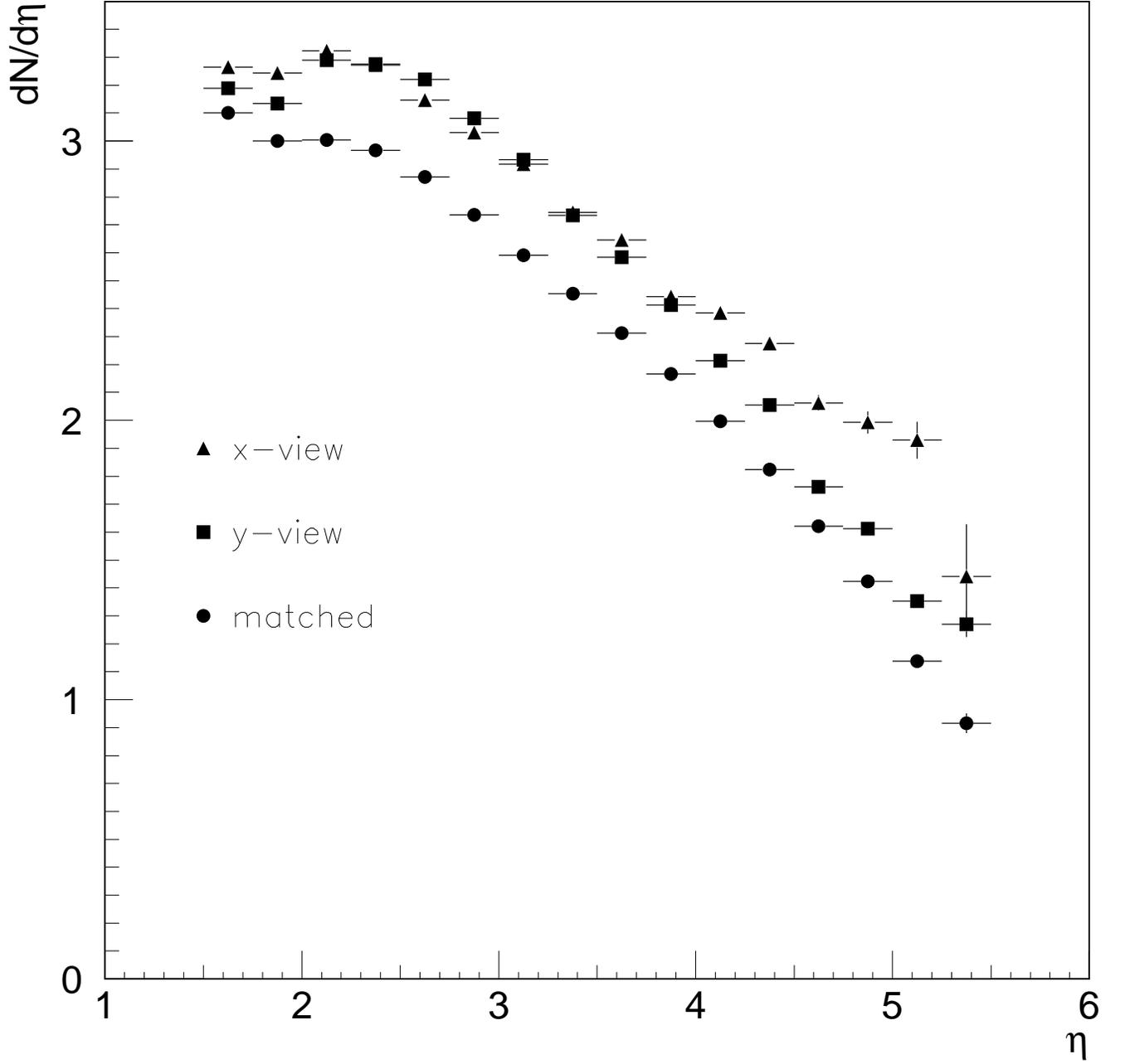}
 \caption{$dN_{ch}/d\eta$ derived from $dN_{ch}/d\eta_x$,
  $dN_{ch}/d\eta_y$, and $dN_{ch}/d\eta_m$.
  The bin width is $0.25$ units of $\eta$.
  Only statistical errors are shown.
 \label{fig:allruns} }
\end{figure}

\begin{figure}
 \epsfclipon \epsfxsize=\figwidth
 \epsffile{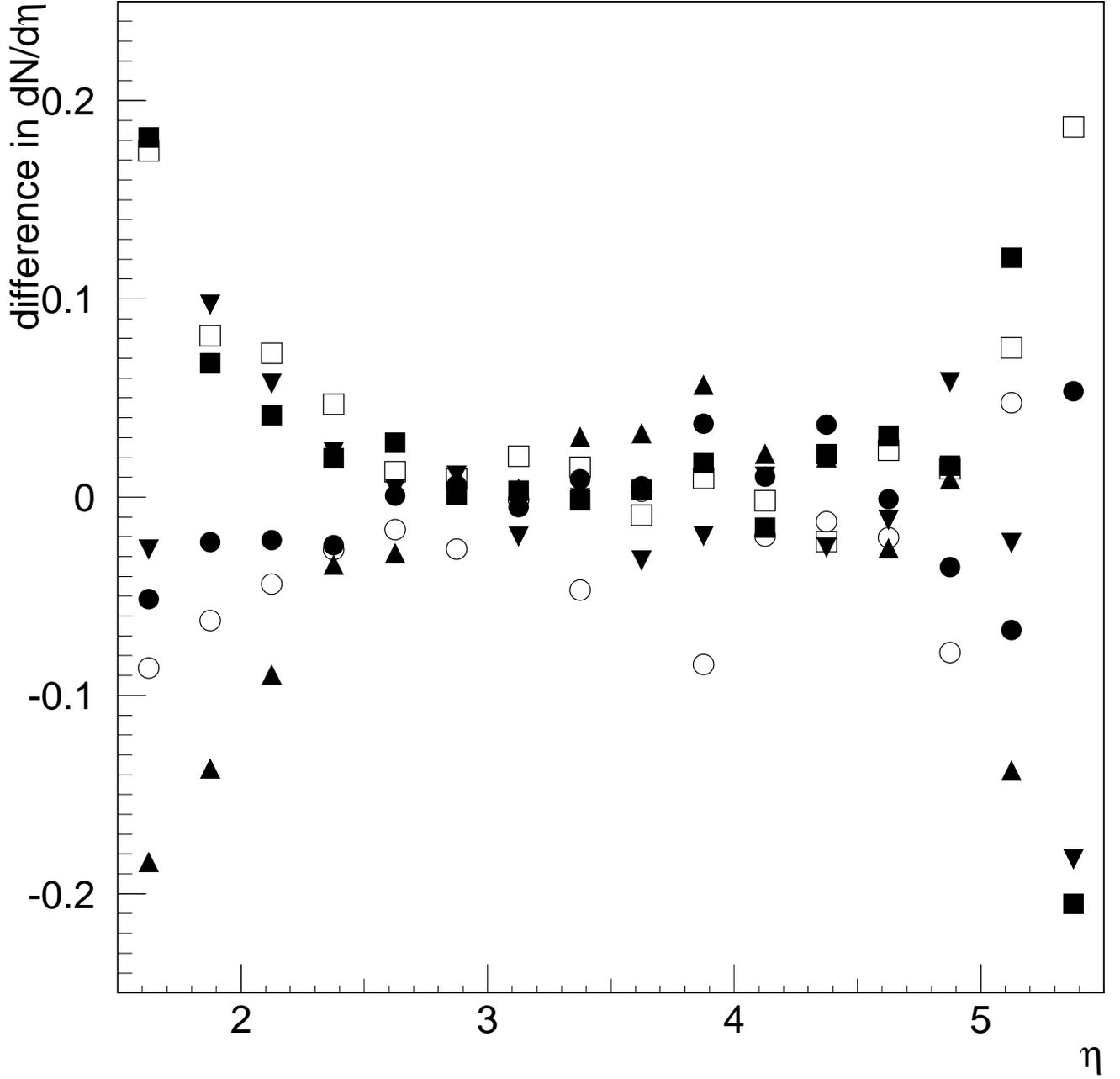}
 \caption{The differences between $dN_{ch}/d\eta$ calculated
  separately for 6 data subsets and the final result.
  The differences arise from a combination of statistical and systematic
  errors and represent part of the shape error quoted in the final result.
 \label{fig:differences} }
\end{figure}

\begin{figure}
 \epsfclipon \epsfxsize=\figwidth
 \epsffile{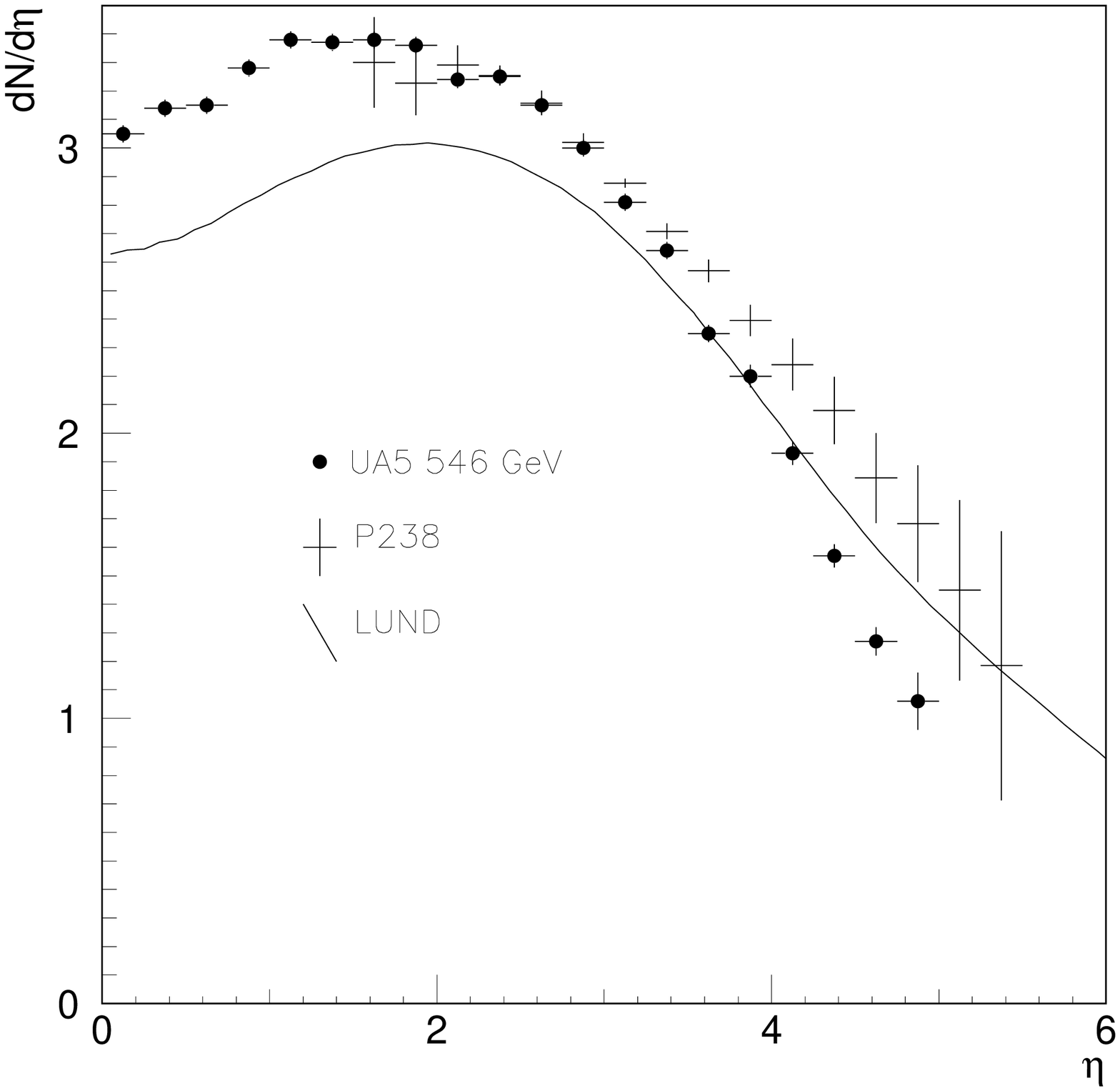}
 \caption{Final $dN_{ch}/d\eta$ result compared to the result from UA5 and
  the distribution generated by PYTHIA-JETSET.
  For the P238 points, only the shape errors are shown, the 5\% normalization
  error is not plotted.
  The UA5 points are plotted with their statistical errors, a normalization
  uncertainty of order 2.5\% is not shown.
  The Monte Carlo curve is unweighted for purposes of comparison.
 \label{fig:result} }
\end{figure}


\begin{references}
\bibitem{pileup}
We start with a gaussian distribution of interactions, close to the
measured distribution of the event sample, then weight the Monte Carlo
events to exactly match the measured distribution.
We enable PYTHIA's
overlayed event option to generate pile-up events, then use event
weightings to fine tune the distributions to match a particular data
set.
\end{references}
\end{document}